\documentclass[usegraphicx]{mn2e}

\title[What powers the starburst activity of NGC 1068?]
      {What powers the starburst activity of NGC 1068?\\
       Star-driven gravitational instabilities caught in the act}

\author[A. B. Romeo and K. Fathi]
       {Alessandro B. Romeo$^{1}$\thanks{E-mail: romeo@chalmers.se}
        and Kambiz Fathi$^{2,3}$\\
        $^{1}$Department of Earth and Space Sciences,
              Chalmers University of Technology,
              SE-41296 Gothenburg, Sweden\\
        $^{2}$Department of Astronomy,
              Stockholm University,
              AlbaNova Centre,
              SE-10691 Stockholm, Sweden\\
        $^{3}$Oskar Klein Centre for Cosmoparticle Physics,
              Stockholm University,
              SE-10691 Stockholm, Sweden}

\begin{document}

\date{Accepted 2016 May 11.
      Received 2016 April 12; in original form 2016 February 09}

\pagerange{\pageref{firstpage}--\pageref{lastpage}}

\pubyear{2016}

\maketitle

\label{firstpage}

\begin{abstract}
We explore the role that gravitational instability plays in NGC 1068, a
nearby Seyfert galaxy that exhibits unusually vigorous starburst activity.
For this purpose, we use the Romeo-Falstad disc instability diagnostics and
data from BIMA SONG, SDSS and SAURON.  Our analysis illustrates that NGC 1068
is a gravitationally unstable `monster'.  Its starburst disc is subject to
unusually powerful instabilities.  Several processes, including AGN/stellar
feedback, try to quench such instabilities from inside out by depressing the
surface density of molecular gas across the central kpc, but they do not
succeed.  Gravitational instability `wins' because it is driven by the stars
via their much higher surface density.  In this process, stars and molecular
gas are strongly coupled, and it is such a coupling that ultimately triggers
local gravitational collapse/fragmentation in the molecular gas.
\end{abstract}


\section{INTRODUCTION}

NGC 1068 is one of the closest and most famous Seyfert 2 galaxies (Khachikian
\& Weedman 1974).  Since the discovery of its Seyfert 1 nucleus in linearly
polarized light (Antonucci \& Miller 1985), it presents the strongest case
for a unified model of active galactic nuclei (AGNs).

\begin{figure}
\includegraphics[scale=.98]{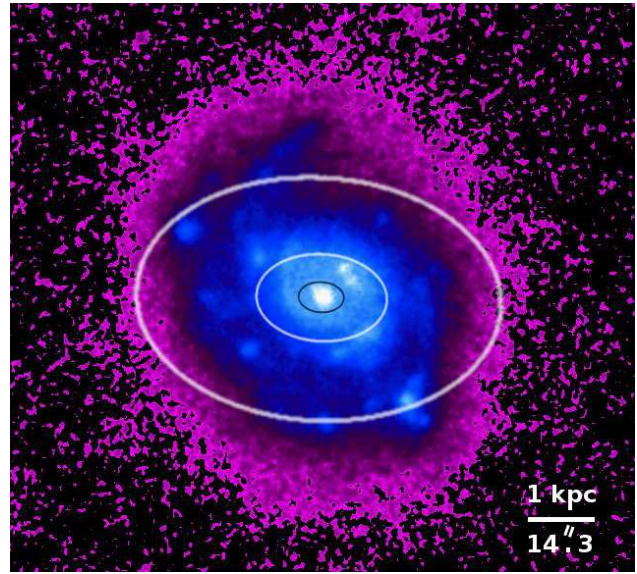}
\caption{False-colour UV image of NGC 1068, from the Ultraviolet Imaging
  Telescope (UIT) on its Astro-1 Space Shuttle mission.  The image is
  displayed in a north-up east-left orientation.  The three ellipses show the
  innermost and the outermost radii of the analysis presented here (350 pc
  and 3 kpc), and the 1 kpc radius.  The ellipses take into account the
  inclination and position angle of the galaxy.  Image courtesy of UIT and
  NASA.}
\end{figure}

NGC 1068 is also a powerful starburst galaxy.  As highlighted by Begelman
(1997), vigorous star formation is observed `out in the open' on kiloparsec
scales (e.g., Lester et al.\ 1987; Telesco \& Decher 1988; Bruhweiler et
al.\ 1991; Hutchings et al.\ 1991; Neff et al.\ 1994).  This is beautifully
illustrated in Fig.\ 1, a false-colour UV image of NGC 1068 observed by the
UIT team (Neff et al.\ 1994; Fanelli et al.\ 1997; Marcum et al.\ 2001).  In
this figure, and in the rest of the paper, we adhere to the `Ringberg
standards' (Bland-Hawthorn et al.\ 1997) and adopt a distance of 14.4 Mpc, so
that 1\arcsec corresponds to 69.8 pc.  Fig.\ 1 shows that, in addition to the
AGN, there is a population of luminous starburst knots spread over a bright
disc of size $R\approx3\,\mbox{kpc}$, which is surrounded by a fainter halo.
The disc emits four times more UV flux than the AGN (Fanelli et al.\ 1997),
and the starburst knots are among the brightest star formation regions known,
perhaps the most luminous within a distance of 30 Mpc (Neff et al.\ 1994).

NGC 1068 displays the classic `bar within bar' morphology, with a starburst
pseudo-ring of radius $R\approx1.1\,\mbox{kpc}$ located at the end of the
secondary bar (see Bland-Hawthorn et al.\ 1997 and references therein).
Schinnerer et al.\ (2000) analysed the structures and rotation curve traced
by the molecular gas, and estimated that the starburst pseudo-ring (appearing
as tightly wound spiral arms in CO) is close both to the inner Lindblad
resonance of the primary bar and to the corotation resonance of the secondary
bar.  Emsellem et al.\ (2006) analysed the gas and stellar dynamics of NGC
1068 in detail, using a variety of data as well as numerical simulations, and
found that there is no clear mode coupling (resonance overlap) between the
two bars.  In particular, the corotation resonance of the secondary bar lies
at $R\approx\mbox{2.2--2.3}\,\mbox{kpc}$, i.e.\ well beyond the end of the
bar ($R\approx1.1\,\mbox{kpc}$).  They also showed that the secondary bar
produces clear signatures in the stellar kinematics, and that this structure
could drive a significant amount of gas down to the central 300 pc (see
Emsellem et al.\ 2006 for more details).

Gravitational instability is expected to play a significant role in this
scenario because of its link with star formation (e.g., Elmegreen 2012; Mac
Low 2013), the growth of bars within bars and associated structures (e.g.,
Shlosman et al.\ 1989).  Brinks et al.\ (1997) and Gallimore et al.\ (1999)
estimated the value of Toomre's (1964) $Q$ stability parameter in the
starburst disc of NGC 1068, considering both atomic and molecular gas, but
they found no clear evidence of gravitational instability.  This result is
not discouraging.  \emph{Disc instabilities are difficult to detect without
  more powerful diagnostics} (Romeo \& Fathi 2015).

Here we explore this important aspect of the problem using the disc
instability diagnostics developed by Romeo \& Wiegert (2011) and Romeo \&
Falstad (2013), which have been used in a variety of contexts (e.g., Hunter
et al.\ 2013; Meurer et al.\ 2013; Zheng et al.\ 2013; Forbes et al.\ 2014;
Genzel et al.\ 2014; Grebovi\'{c} 2014; Romeo \& Agertz 2014; Westfall et
al.\ 2014; Yim et al.\ 2014; Agertz et al.\ 2015; Fathi et al.\ 2015;
Goldbaum et al.\ 2015; Romeo \& Fathi 2015; Inoue et al.\ 2016; Khoperskov et
al.\ 2016; Williamson et al.\ 2016).  To detect gravitational instabilities
across the starburst disc of NGC 1068, we consider not only molecular gas but
also a component that is still often disregarded when analysing the stability
of spiral galaxies: the stars!  We do not consider atomic gas because, in the
starburst disc of NGC 1068, it has much lower surface density and higher
velocity dispersion than molecular gas (Brinks et al.\ 1997; Gallimore et
al.\ 1999), hence it contributes much less to disc instability (Romeo \&
Falstad 2013).  For a similar reason, we do not consider ionized gas.  The
data and method are described in Sect.\ 2, the results are presented in
Sect.\ 3 and discussed in Sect.\ 4, and the conclusions are drawn in
Sect.\ 5.

\section{DATA AND METHOD}

As motivated in Sect.\ 1, we consider molecular hydrogen gas, as traced by CO
emission, and stars.  A proper two-component stability analysis requires
radial profiles of five basic quantities: the epicyclic frequency, $\kappa$,
the surface densities of molecular gas and stars, $\Sigma_{\mathrm{co}}$ and
$\Sigma_{\star}$, the molecular 1D (line-of-sight) velocity dispersion,
$\sigma_{\mathrm{co}}$, and the stellar radial velocity dispersion,
$\sigma_{\star}$.  Although for simplicity we use similar notations, the last
two quantities reflect an important dynamical difference between molecular
gas and stars.  To first approximation, molecular gas is collisional so its
velocity dispersion is isotropic, while the stellar component is
collisionless and has an anisotropic velocity dispersion.  We derive
$\Sigma_{\mathrm{co}}(R)$, $\sigma_{\mathrm{co}}(R)$ and $\kappa(R)$ in
Sect.\ 2.1, and $\Sigma_{\star}(R)$ and $\sigma_{\star}(R)$ in Sect.\ 2.2.
Finally, we present the disc instability diagnostics and compute their radial
profiles in Sect.\ 2.3.

\subsection{Molecular gas}

We use CO $J(1\to0)$ data from the BIMA Survey of Nearby Galaxies (BIMA SONG;
Helfer et al.\ 2003), which have a spatial resolution of
$8.9\arcsec\times5.6\arcsec$, sampled at 1\arcsec per pixel, and a velocity
resolution of 10 km\,s$^{-1}$.  To derive $\Sigma_{\mathrm{co}}(R)$,
$\sigma_{\mathrm{co}}(R)$ and $\kappa(R)$ in NGC 1068, we use the same
methodology as for NGC 6946 (Romeo \& Fathi 2015) and NGC 7469 (Fathi et
al.\ 2015), which we re-describe below.  Each part of the following procedure
was thoroughly tested by Romeo \& Fathi (2015).

We derive spatial maps of $\Sigma_{\mathrm{co}}$ and $\sigma_{\mathrm{co}}$
by applying Gaussian fits to the individual spectra.  We find that our
amplitude map matches the zeroth-moment map presented by the BIMA SONG team.
The agreement is within 20\%, except where the CO flux approaches the
detection limit of the BIMA SONG survey ($R\ga2\,\mbox{kpc}$).  One advantage
of applying Gaussian fits is that we can get better covering maps if we use a
Hanning smoothing algorithm in the spectral direction before fitting the
individual Gaussian profiles (e.g., Hernandez et al.\ 2005; Daigle et
al.\ 2006).  This procedure does not affect the line amplitude or shift.
However, it artificially broadens the individual spectra by
$\delta\sigma_{\mathrm{co}}\approx4.4\;\mbox{km\,s}^{-1}$, which we then
subtract quadratically from the derived velocity dispersion map.  Finally, we
apply a cleaning procedure by removing all the spectra for which the emission
line amplitude is smaller than twice the rms noise (21\% of the spectra), or
$\sigma_{\mathrm{co}}$ is smaller than half the velocity channel (3\% of the
spectra), or the formal error in the derived $\sigma_{\mathrm{co}}$ is
greater than 10 km\,s$^{-1}$ (5\% of the spectra), where the percentage
specified in each case is computed inside a box of 6 kpc ($R=3\,\mbox{kpc}$).
In total, 24\% of the spectra are removed.  These are pixels where the CO
flux approaches the detection limit of the BIMA SONG survey
($R\ga2\,\mbox{kpc}$).  Therefore our cleaning procedure does not introduce
any significant physical bias in the analysis of molecular gas.  A careful
inspection of the individual spectra reveals the presence of multiple
components at different locations, mostly in the central few 100 pc.  In view
of the good agreement between our amplitude map and the zeroth-moment map
presented by the BIMA SONG team, and in view of the complications involved in
multiple-profile fitting schemes (Blasco-Herrera et al.\ 2010), we do not use
profile decomposition methods.  All subsequent analysis is therefore based on
single-Gaussian profile fitting.

We derive $\Sigma_{\mathrm{co}}(R)$ and $\sigma_{\mathrm{co}}(R)$ using
robust statistics, which are especially useful when the data are few or
contain a significant fraction of outliers, or even when the data deviate
significantly from a normal distribution (e.g., Rousseeuw 1991; M\"{u}ller
2000; Romeo et al.\ 2004; Huber \& Ronchetti 2009; Feigelson \& Babu 2012).
To derive the radial profiles of $\Sigma_{\mathrm{co}}$ and
$\sigma_{\mathrm{co}}$, we divide their maps into tilted rings, which are
circular in the plane of the galaxy, and compute the median values of
$\Sigma_{\mathrm{co}}$ and $\sigma_{\mathrm{co}}$ in each ring.  We then
estimate the uncertainty in these median values via the median absolute
deviation (MAD):
\begin{equation}
\Delta X_{\mathrm{med}}=1.858\times\mbox{MAD}/\sqrt{N}\,,
\end{equation}
\begin{equation}
\mbox{MAD}=\mbox{median}\{|X_{i}-X_{\mathrm{med}}|\}\,,
\end{equation}
where $X_{i}$ are the individual measurements of $\Sigma_{\mathrm{co}}$ or
$\sigma_{\mathrm{co}}$, $X_{\mathrm{med}}$ is their median value, and $N$ is
the number of resolution elements in each ring (i.e.\ the number of pixels in
the ring divided by the number of pixels per resolution element).  Eqs (1)
and (2) are the robust counterparts of the formula traditionally used for
estimating the uncertainty in the mean: $\Delta
X_{\mathrm{mean}}=\mbox{SD}/\sqrt{N}$, where SD denotes the standard
deviation (see again M\"{u}ller 2000).%
\footnote{In Eq.\ (1), the numerical constant $C\simeq1.858$ is the product
  of two factors: $C=C_{1}C_{2}$.  The factor $C_{1}\simeq1/0.6745$ converts
  the median absolute deviation into a robust estimate of standard deviation:
  $\mbox{SD}=C_{1}\,\mbox{MAD}$ [see eq.\ (6) and following unnumbered
    equation of M\"{u}ller 2000].  The factor $C_{2}=\sqrt{\pi/2}\simeq1.253$
  converts uncertainty in the mean into uncertainty in the median: $\Delta
  X_{\mathrm{med}}=C_{2}\,\Delta X_{\mathrm{mean}}=C_{2}\,\mbox{SD}/\sqrt{N}$
  [see eq.\ (8) of M\"{u}ller 2000].}
Indeed, the median and the median absolute deviation provide robust
statistical estimates of the `central value' and the `width' of a data set,
respectively, even when almost 50\% of the data are outliers!  Our numerous
tests with varying ring widths and radii confirm that the high quality of the
BIMA SONG data allows a derivation of $\Sigma_{\mathrm{co}}$ and
$\sigma_{\mathrm{co}}$ in rings narrower than the synthesized beam size of
the interferometric observations.  This is mainly thanks to the good sampling
of the resolution element.  The smallest reliable step is found to be 200 pc.
This is the step size adopted throughout our analysis.  A step size of 300 pc
does not change the derived values of $\Sigma_{\mathrm{co}}$ and
$\sigma_{\mathrm{co}}$ significantly, but makes their profiles sparsely
sampled.  As motivated in Sect.\ 1, the $\Sigma_{\mathrm{co}}$ and
$\sigma_{\mathrm{co}}$ profiles presented here are derived up to
$R=3\,\mbox{kpc}$, but we exclude the central $R=350\,\mbox{pc}$ because such
a region is highly disturbed by a massive AGN-driven outflow
(Garc\'{i}a-Burillo et al.\ 2014).  The surface density of molecular gas is
converted to physical units by adopting the standard CO-to-$\mathrm{H}_{2}$
conversion factor,
$X_{\mathrm{co}}=2\times10^{20}\;\mbox{cm}^{-2}\;(\mbox{K\,km\,s}^{-1})^{-1}$
(Bolatto et al.\ 2013), consistent with Helfer et al.\ (2003).  We also
correct for the contribution of helium multiplying by a factor of 1.36.

We derive $\kappa(R)$ from the observed rotation curve.  To calculate the
rotation curve, we assume that circular rotation is the dominant kinematic
feature, and that our measurements refer to positions on a single inclined
disc.  We then use the tilted ring method combined with the harmonic
decomposition formalism (e.g., Schoenmakers et al.\ 1997; Wong et al.\ 2004;
Fathi et al.\ 2005).  This procedure does not correct for beam smearing
(e.g., Schinnerer et al.\ 2000; Teuben 2002; Cald\'{u}-Primo et al.\ 2013).
However, it corrects for and automatically models departures from circular
motion, which are significant in NGC 1068 (Fathi 2004; Emsellem et
al.\ 2006).  Given that this procedure involves fitting several parameters at
each radius of the observed velocity field, contrary to simply finding the
median values of $\Sigma_{\mathrm{co}}$ and $\sigma_{\mathrm{co}}$, we cannot
use the same initial step size.  Hence we apply a larger radial step, and
interpolate linearly to obtain the rotation curve at all radii where we have
calculated the robust $\Sigma_{\mathrm{co}}$ and $\sigma_{\mathrm{co}}$
values.  Once the rotation curve $V_{\mathrm{rot}}$ is calculated at each
radius $R$, we derive the angular frequency $\Omega=V_{\mathrm{rot}}/R$ and
the epicyclic frequency $\kappa=\sqrt{R\,d\Omega^{2}/dR+4\Omega^{2}}$.

\subsection{Stars}

We adopt the high-quality $\Sigma_{\star}(R)$ derived by Bakos \& Trujillo
(2012) from Stripe 82 of the Sloan Digital Sky Survey (SDSS; Abazajian et
al.\ 2009).  We also adopt the high-quality spatial map of stellar
line-of-sight velocity dispersion ($\sigma_{\mathrm{los}\star}$) derived by
Fathi (2004) and Emsellem et al.\ (2006) from observations with the
Spectrographic Areal Unit for Research on Optical Nebulae (SAURON; Bacon et
al.\ 2001).

To derive $\sigma_{\star}(R)$ from the SAURON $\sigma_{\mathrm{los}\star}$
map, we use the following method.  We start from a general formula that
expresses the line-of-sight velocity dispersion ($\sigma_{\mathrm{los}}$) in
terms of the radial ($\sigma_{R}$), tangential ($\sigma_{\phi}$) and vertical
($\sigma_{z}$) components:
\begin{equation}
\sigma_{\mathrm{los}}^{2}=
\left(\sigma_{R}^{2}\sin^{2}\phi+\sigma_{\phi}^{2}\cos^{2}\phi\right)\sin^{2}i+
\sigma_{z}^{2}\cos^{2}i\,,
\end{equation}
where $i$ is the inclination angle of the galaxy, and $\phi$ is the position
angle with respect to the major axis in the plane of the disc (see, e.g.,
Binney \& Merrifield 1998).  We introduce the two parameters
$A=\sigma_{\phi}/\sigma_{R}$ and $B=\sigma_{z}/\sigma_{R}$, and rewrite
Eq.\ (3) as
\begin{equation}
\sigma_{R}=\sigma_{\mathrm{los}}
\left[\left(\sin^{2}\phi+A^{2}\cos^{2}\phi\right)\sin^{2}i+
      B^{2}\cos^{2}i\right]^{-1/2}\,.
\end{equation}
While $A_{\mathrm{co}}\approx B_{\mathrm{co}}\approx1$ (see Sect.\ 2), both
$A_{\star}$ and $B_{\star}$ are significantly different from unity.  In fact,
from the epicyclic approximation it follows that
$A_{\star}\approx\kappa/2\Omega$ (see, e.g., Binney \& Tremaine 2008).  And,
for an Sb galaxy like NGC 1068, $B_{\star}\approx0.6$ (Shapiro et al.\ 2003;
Gerssen \& Shapiro Griffin 2012).  We estimate $A_{\star}$ and $B_{\star}$ as
above, and compute the spatial map of $\sigma_{\star}$ from the SAURON
$\sigma_{\mathrm{los}\star}$ map via Eq.\ (4).  Finally, we derive
$\sigma_{\star}(R)$ using the same method as for $\sigma_{\mathrm{co}}(R)$,
but only up to $R=2\,\mbox{kpc}$.  Our numerous tests with varying ring radii
and widths show that this is the outer limit for extracting reliable stellar
velocity dispersions from the SAURON data.

\subsection{Disc instability diagnostics}

We use two disc instability diagnostics derived by Romeo \& Falstad (2013):
\begin{enumerate}
\item a simple and accurate approximation for the $Q$ stability parameter in
  multi-component and realistically thick discs;
\item a corresponding approximation for the characteristic instability
  wavelength, i.e.\ the scale at which the disc becomes locally unstable as
  $Q$ drops below unity.
\end{enumerate}
In our case (molecular gas plus stars) the $Q$ stability parameter,
$\mathcal{Q}_{\mathrm{co}+\star}$, and the characteristic instability
wavelength, $\lambda_{\mathrm{co}+\star}$, are given by
\begin{equation}
\frac{1}{\mathcal{Q}_{\mathrm{co}+\star}}=
\left\{\begin{array}{ll}
       {\displaystyle\frac{1}{T_{\mathrm{co}}Q_{\mathrm{co}}}+
                     \frac{W}{T_{\star}Q_{\star}}}
                       & \mbox{if\ \ }T_{\mathrm{co}}Q_{\mathrm{co}}\leq
                                      T_{\star}Q_{\star}\,,             \\
                       &                                                \\
       {\displaystyle\frac{W}{T_{\mathrm{co}}Q_{\mathrm{co}}}+
                     \frac{1}{T_{\star}Q_{\star}}}
                       & \mbox{if\ \ }T_{\mathrm{co}}Q_{\mathrm{co}}\geq
                                      T_{\star}Q_{\star}\,,
       \end{array}
\right.
\end{equation}
\mbox{}
\begin{equation}
\lambda_{\mathrm{co}+\star}=
\left\{\begin{array}{ll}
       {\displaystyle2\pi\,\frac{\sigma_{\mathrm{co}}}{\kappa}}
                       & \mbox{if\ \ }T_{\mathrm{co}}Q_{\mathrm{co}}<
                                      T_{\star}Q_{\star}\,,          \\
                       &                                             \\
       {\displaystyle2\pi\,\frac{\sigma_{\star}}{\kappa}}
                       & \mbox{if\ \ }T_{\mathrm{co}}Q_{\mathrm{co}}>
                                      T_{\star}Q_{\star}\,,
       \end{array}
\right.
\end{equation}
where $Q_{i}=\kappa\sigma_{i}/\pi G\Sigma_{i}$ is the Toomre parameter of
component $i$, $\sigma$ denotes the radial velocity dispersion, and $T_{i}$
and $W$ are given by
\begin{equation}
T_{i}=
\left\{\begin{array}{ll}
       {\displaystyle1+0.6\left(\frac{\sigma_{z}}{\sigma_{R}}\right)_{i}^{2}}
                       & \mbox{if\ }0\leq(\sigma_{z}/\sigma_{R})_{i}\leq0.5\,,
                         \\
                       & \\
       {\displaystyle0.8+0.7\left(\frac{\sigma_{z}}{\sigma_{R}}\right)_{i}}
                       & \mbox{if\ }0.5\leq(\sigma_{z}/\sigma_{R})_{i}\leq1\,,
       \end{array}
\right.
\end{equation}
\mbox{}
\begin{equation}
W=
\frac{2\sigma_{\mathrm{co}}\sigma_{\star}}
     {\sigma_{\mathrm{co}}^{2}+\sigma_{\star}^{2}}\,.
\end{equation}
This set of equations tells us that the values of
$\mathcal{Q}_{\mathrm{co}+\star}$ and $\lambda_{\mathrm{co}+\star}$ are
controlled by the component with smaller $TQ$.  \emph{This is the component
  that drives disc instabilities} if $\mathcal{Q}_{\mathrm{co}+\star}\leq1$.
The contribution of the other component is weakened by the $W$ factor.  For
example, if disc instabilities are driven by molecular gas \emph{and}
$\sigma_{\mathrm{co}}\ll\sigma_{\star}$, then (to leading order)
$\mathcal{Q}_{\mathrm{co}+\star}\approx T_{\mathrm{co}}Q_{\mathrm{co}}$ and
$\lambda_{\mathrm{co}+\star}=2\pi\sigma_{\mathrm{co}}/\kappa$.  Vice versa,
in the case of star-driven instabilities,
$\mathcal{Q}_{\mathrm{co}+\star}\approx T_{\star}Q_{\star}$ and
$\lambda_{\mathrm{co}+\star}=2\pi\sigma_{\star}/\kappa$.  The radial profiles
of $\mathcal{Q}_{\mathrm{co}+\star}$ and $\lambda_{\mathrm{co}+\star}$ can
easily be computed from the radial profiles derived in Sects 2.1 and 2.2, and
from the values of $T_{\mathrm{co}}$ and $T_{\star}$.  If molecular gas is
collisional, as is generally assumed, then
$(\sigma_{z}/\sigma_{R})_{\mathrm{co}}\approx1$ and
$T_{\mathrm{co}}\approx1.5$.  In contrast, the velocity dispersion anisotropy
of stars in NGC 1068 is $(\sigma_{z}/\sigma_{R})_{\star}\approx0.6$ (Shapiro
et al.\ 2003), hence $T_{\star}\approx1.2$.

To illustrate the predictive power of a proper two-component stability
analysis, we will also consider molecular gas alone.  In this simple case,
the $Q$ stability parameter and the characteristic instability wavelength are
$\mathcal{Q}_{\mathrm{co}}=\frac{3}{2}\,(\kappa\sigma_{\mathrm{co}}/\pi
G\Sigma_{\mathrm{co}})$ and
$\lambda_{\mathrm{co}}=2\pi\sigma_{\mathrm{co}}/\kappa$.

\section{RESULTS}

\subsection{Molecular gas}

Let us first consider molecular gas, i.e.\ the component that is generally
regarded as the main driver of disc instabilities in the inner regions of
spiral galaxies (e.g., Kennicutt 1989; Martin \& Kennicutt 2001; Romeo \&
Falstad 2013).

\begin{figure}
\includegraphics[angle=-90.,scale=.95]{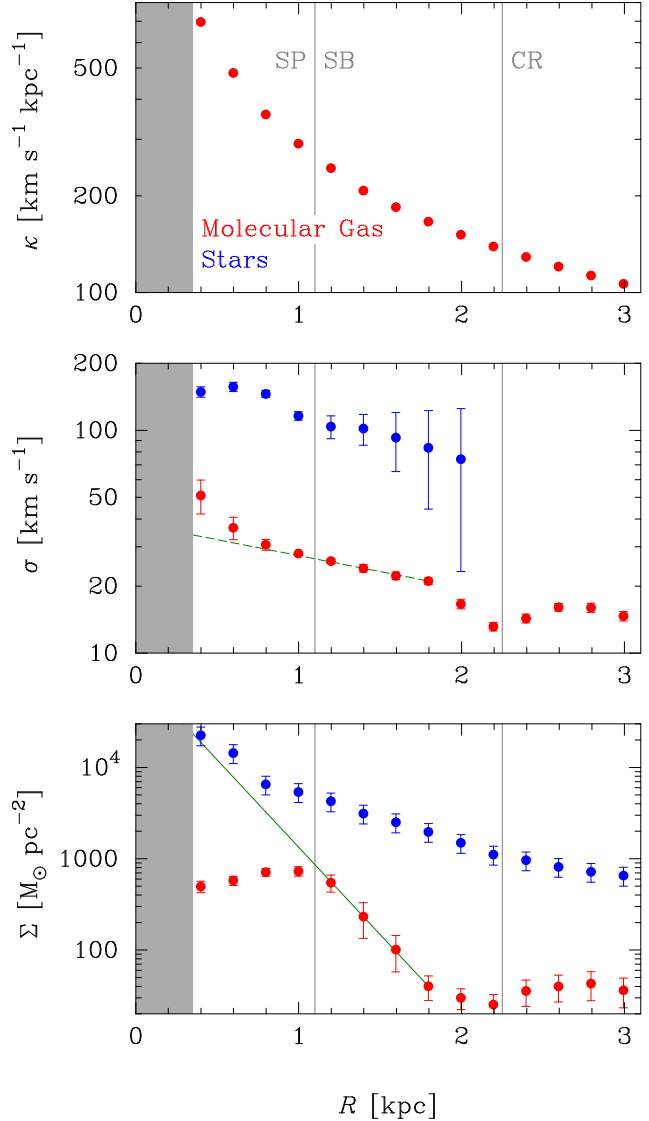}
\caption{Radial profiles of the epicyclic frequency (top), radial velocity
  dispersion (middle) and surface density (bottom) of molecular gas and stars
  in NGC 1068.  The vertical lines represent the approximate radii of the
  starburst pseudo-ring (SP), secondary bar (SB) and its corotation resonance
  (CR).  The bottom and middle panels also show robust, median-based,
  exponential fits to $\Sigma_{\mathrm{co}}(R)$ and $\sigma_{\mathrm{co}}(R)$
  for $1\,\mbox{kpc}<R<2\,\mbox{kpc}$, extrapolated into the central kpc.
  The shaded region is highly disturbed by a massive AGN-driven outflow
  (Garc\'{i}a-Burillo et al.\ 2014), so the following central
  ($R=200\,\mbox{pc}$) values are not shown and are excluded from our
  analysis: $\kappa=860\;\mbox{km\,s}^{-1}\,\mbox{kpc}^{-1}$,
  $\sigma_{\mathrm{co}}=80\pm30\;\mbox{km\,s}^{-1}$ and
  $\sigma_{\star}=130\pm10\;\mbox{km\,s}^{-1}$,
  $\Sigma_{\mathrm{co}}=(5.7\pm0.8)\times10^{2}\;
  \mbox{M}_{\odot}\,\mbox{pc}^{-2}$ and
  $\Sigma_{\star}=(3.1\pm0.7)\times10^{4}\;
  \mbox{M}_{\odot}\,\mbox{pc}^{-2}$.}
\end{figure}

Fig.\ 2 shows that $\Sigma_{\mathrm{co}}(R)$ is depressed at small radii
relative to an exponential radial distribution, as was also pointed out by
Regan et al.\ (2001) and Helfer et al.\ (2003).  Exponential distributions
are traditionally used for fitting the radial profiles of CO surface
brightness observed in nearby galaxies, and thus for inferring important
physical parameters such as the scale length of the molecular disc and, when
relevant, its central surface density (e.g., Young \& Scoville 1982; Young et
al.\ 1995; Sakamoto et al.\ 1999; Regan et al.\ 2001; Helfer et al.\ 2003;
Leroy et al.\ 2008, 2009).  Fig.\ 2 shows that $\Sigma_{\mathrm{co}}(R)$ is
well fitted by an exponential distribution only within a narrow radial range,
$1\,\mbox{kpc}<R<2\,\mbox{kpc}$, i.e.\ approximately between the starburst
pseudo-ring ($R\approx1.1\,\mbox{kpc}$) and the corotation resonance of the
secondary bar ($R\approx\mbox{2.2--2.3}\,\mbox{kpc}$).  This makes sense
because the central kpc is disturbed not only by a massive AGN-driven outflow
(Garc\'{i}a-Burillo et al.\ 2014) but also by strong inward streaming motions
driven by the starburst pseudo-ring (Emsellem et al.\ 2006), while radii
$R\ga2\,\mbox{kpc}$ are disturbed by a change in flow across the corotation
resonance of the secondary bar.  This is particularly well illustrated by the
presence of a cusp in the radial profile of $\sigma_{\mathrm{co}}$.  In
addition, for $R\ga2\,\mbox{kpc}$ the CO flux approaches the detection limit
of the BIMA SONG survey (Regan et al.\ 2001; Helfer et al.\ 2003).

A robust, median-based, exponential fit to $\Sigma_{\mathrm{co}}(R)$ for
$1\,\mbox{kpc}<R<2\,\mbox{kpc}$ yields a scale length of about 230 pc and an
extrapolated value of $\Sigma_{\mathrm{co}}$ at $R=350\,\mbox{pc}$ of about
$2\times10^{4}\;\mbox{M}_{\odot}\,\mbox{pc}^{-2}$, comparable to the stellar
value.  Given the small number of data points in this radial range, and given
their non-negligible error bars, it might be that even a robust fit is
significantly biased.  However, the resulting scale length makes sense
because it is similar to the scale length that characterizes the nuclear
molecular disc of another bar-within-bar galaxy: NGC 6946 (Romeo \& Fathi
2015).  The central $\Sigma_{\mathrm{co}}(R)$ depression has instead no match
in NGC 6946.  For a galaxy with a massive AGN-driven outflow and intense
starburst activity such as NGC 1068, it seems natural to associate the
central $\Sigma_{\mathrm{co}}(R)$ depression with AGN/stellar feedback (e.g.,
Karouzos et al.\ 2014; Storchi-Bergmann 2014; Agertz \& Kravtsov 2015;
Melioli \& de Gouveia Dal Pino 2015; Garc\'{i}a-Burillo 2016; Hopkins et
al.\ 2016).  However, as pointed out by the referee, other processes can
contribute to the central $\Sigma_{\mathrm{co}}(R)$ depression, e.g.\ the
consumption of molecular gas by star formation and the transport of molecular
gas inwards by the action of the bar(s).

As can be seen in Fig.\ 2, $\sigma_{\mathrm{co}}(R)$ also shows an
exponential fall-off for $1\,\mbox{kpc}<R<2\,\mbox{kpc}$, so we have also
robustly fitted $\sigma_{\mathrm{co}}(R)$ in this radial range and
extrapolated the fit back into the central kpc.  The resulting scale length
is about 3 kpc and the extrapolated value of $\sigma_{\mathrm{co}}$ at
$R=350\,\mbox{pc}$ is about 35 km\,s$^{-1}$.  Note, however, that the rise of
$\Delta\sigma_{\mathrm{co}}(R)=\sigma_{\mathrm{co}}(R)-\sigma_{\mathrm{fit}}(R)$
towards the centre can be caused by other factors than the processes
mentioned above, e.g.\ disc heating (Kormendy \& Kennicutt 2004; Romeo \&
Fathi 2015) and beam smearing (Schinnerer et al.\ 2000; Teuben 2002;
Cald\'{u}-Primo et al.\ 2013).

\subsection{Is the starburst disc gravitationally unstable?}

\begin{figure}
\includegraphics[angle=-90.,scale=.96]{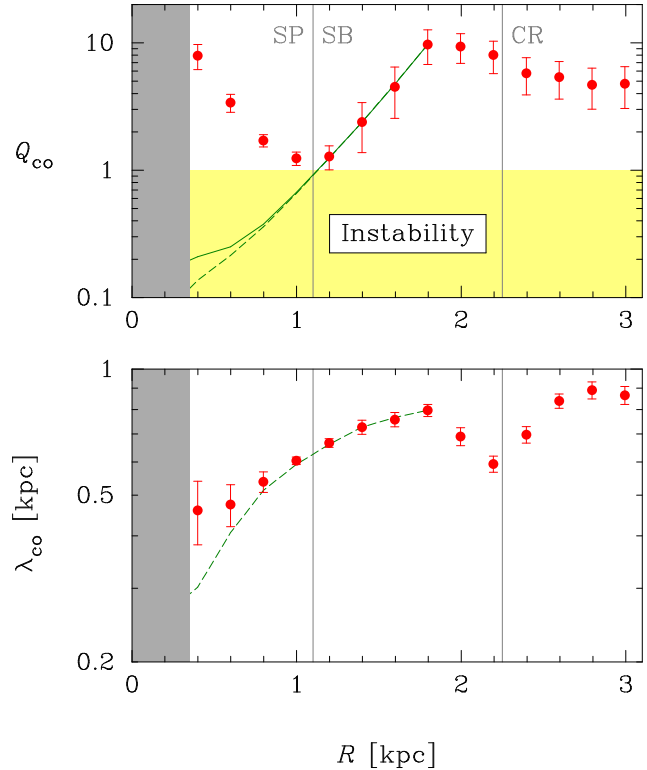}
\caption{Radial profiles of the one-component $\mathcal{Q}$ stability
  parameter (top) and characteristic instability wavelength (bottom) for the
  disc of molecular gas in NGC 1068.  The vertical lines represent the
  approximate radii of the starburst pseudo-ring (SP), secondary bar (SB) and
  its corotation resonance (CR).  Also shown are predictions based on the fit
  to $\Sigma_{\mathrm{co}}(R)$ (solid line) and the fits to both
  $\Sigma_{\mathrm{co}}(R)$ and $\sigma_{\mathrm{co}}(R)$ (dashed lines)
  displayed in Fig.\ 2.}
\end{figure}

Fig.\ 3 shows that $\mathcal{Q}_{\mathrm{co}}(R)$ has a deep minimum at
$R\approx1.1\,\mbox{kpc}$, where its value gets close to unity, while the
radial profile of $\mathcal{Q}_{\mathrm{co}}$ extrapolated using
$\Sigma_{\mathrm{fit}}(R)$ (solid line) is well below unity for
$R<1\,\mbox{kpc}$.  Therefore the effect of the central
$\Sigma_{\mathrm{co}}(R)$ depression is to raise the
$\mathcal{Q}_{\mathrm{co}}$ stability parameter beyond its critical value
(compare the data points with the solid line).  This quenches molecular-gas
instabilities from inside out, leaving the molecular disc marginally unstable
only within an extremely narrow radial range: the starburst pseudo-ring.
Fig.\ 3 also shows that the radial profile of $\mathcal{Q}_{\mathrm{co}}$
extrapolated using both $\Sigma_{\mathrm{fit}}(R)$ and
$\sigma_{\mathrm{fit}}(R)$ (dashed line) is lower than the one extrapolated
using only $\Sigma_{\mathrm{fit}}(R)$ (solid line) for $R<1\,\mbox{kpc}$.  So
the rise of $\Delta\sigma_{\mathrm{co}}(R)$ towards the centre strengthens
the stabilizing effect of the central $\Sigma_{\mathrm{co}}(R)$ depression.
And it does so by diluting molecular-gas instabilities over larger scales
(see now the bottom panel of Fig.\ 3, and compare the data points with the
dashed line).  Last but not least, note that $\lambda_{\mathrm{co}}(R)$ has a
clear drop at $R\approx\mbox{2.2--2.3}\,\mbox{kpc}$.  This result confirms
that the radial profile of the characteristic instability wavelength is a
powerful diagnostic for predicting the corotation resonances of bars within
bars (Romeo \& Fathi 2015).

\subsection{Don't forget the stars!}

\begin{figure}
\includegraphics[angle=-90.,scale=.95]{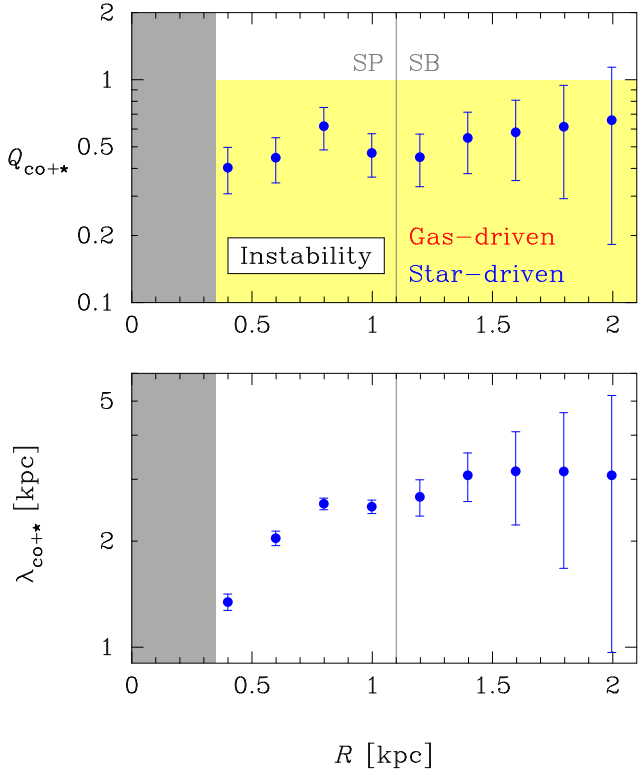}
\caption{Radial profiles of the two-component $\mathcal{Q}$ stability
  parameter (top) and characteristic instability wavelength (bottom) for the
  disc of molecular gas and stars in NGC 1068.  The data are colour-coded so
  as to show whether disc instabilities are driven by molecular gas or stars.
  The vertical line represents the approximate radius of the starburst
  pseudo-ring (SP) and secondary bar (SB).}
\end{figure}

Are gravitational instabilities really so marginally important in the
starburst disc of NGC 1068?  And does the central $\Sigma_{\mathrm{co}}(R)$
depression really have such a strong impact on them?  To answer these
questions, we \emph{must} consider a component that is still often
disregarded when analysing the stability of spiral galaxies: the stars.
$\Sigma_{\star}(R)$ and $\sigma_{\star}(R)$ are shown in Fig.\ 2, while the
resulting $\mathcal{Q}_{\mathrm{co}+\star}(R)$ and
$\lambda_{\mathrm{co}+\star}(R)$ are shown in Fig.\ 4.  This figure
illustrates two important results:
\begin{enumerate}
\item the starburst disc of NGC 1068 is subject to violent and large-scale
  gravitational instabilities;
\item in the current evolutionary phase of NGC 1068, the central
  $\Sigma_{\mathrm{co}}(R)$ depression does not have a strong impact on disc
  instabilities because these are entirely driven by the stars.
\end{enumerate}
At $R=2\,\mbox{kpc}$, the outer radius imposed by the SAURON data, there is
still no sign of transition to local stability or to a gas-driven regime.
This is unusual for a nearby star-forming galaxy, as we will discuss in
Sect.\ 4 together with other aspects of the problem.

\begin{figure}
\includegraphics[scale=1.]{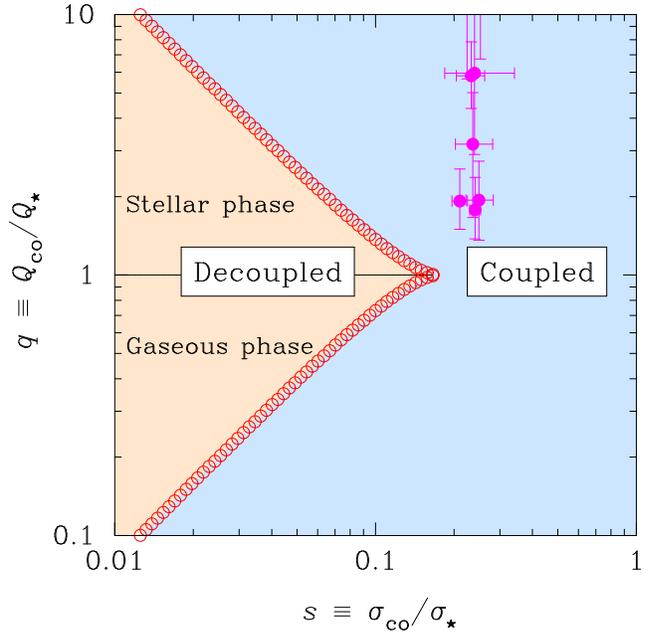}
\caption{The parameter plane of two-component disc instabilities populated by
  the NGC 1068 data.  Here $\sigma_{\mathrm{co}}$ and $\sigma_{\star}$ are
  the radial velocity dispersions of molecular gas and stars,
  $Q_{\mathrm{co}}$ and $Q_{\star}$ are their Toomre parameters.  Outside the
  `two-phase region', the responses of the two components to perturbations
  are coupled.  So star-driven instabilities can also lead to local
  gravitational collapse/fragmentation in the molecular gas.}
\end{figure}

Can star-driven instabilities lead to local gravitational
collapse/fragmentation in the molecular gas?  To answer this question, we
should understand in more detail how molecular gas and stars contribute to
disc instabilities.  This is shown in Fig.\ 5.  Inside the `two-phase
region', the contributions of molecular gas and stars to the gravitational
instability of the disc peak at two different wavelengths (Bertin \& Romeo
1988; Romeo \& Wiegert 2011).  In the `gaseous phase', the short-wavelength
peak is higher than the other one and molecular gas will dominate the onset
of gravitational instability.  Vice versa, in the `stellar phase', the
long-wavelength peak is higher and stars will dominate.  The shape and size
of this region are only moderately affected by disc thickness (Romeo \&
Wiegert 2011), gas turbulence (Hoffmann \& Romeo 2012) or the fact that the
stellar component is collisionless (Romeo \& Falstad 2013).  In the rest of
the parameter plane, the dynamical responses of the two components are
strongly coupled and peak at a single wavelength.  This means that any
instability driven by one of the components will also perturb and destabilize
the other.  In particular, star-driven instabilities will lead to local
gravitational collapse/fragmentation in the molecular gas.  This is clearly
the case for NGC 1068, as all of the data fall outside the two-phase region.

\section{DISCUSSION}

\subsection{How unusual is NGC 1068?}

To estimate how unusual NGC 1068 is, consider the nearby star-forming
galaxies that have been analysed using the Romeo-Falstad disc instability
diagnostics (see references in Sect.\ 1).  These are more than 50 galaxies
selected from BIMA SONG, DiskMass, HERACLES, SINGS, THINGS and other surveys.
Among these galaxies, NGC 1068 is the only one in which $\mathcal{Q}(R)$ is
well below unity across such a wide radial range (see Fig.\ 4).  NGC 1068 is
also the only galaxy in which disc instabilities are entirely driven by the
stars.  Most of the other galaxies are instead in \emph{stable}
($\langle\mathcal{Q}\rangle\sim2$) star-driven regimes characterized by a
strong star-gas coupling.  For THINGS spirals, this is illustrated in figs
3--5 of Romeo \& Falstad (2013).  In view of these facts, it seems natural to
associate the unusually violent and large-scale gravitational instabilities
detected in NGC 1068 with the unusually intense and widespread starbursts
observed in this galaxy (see references in Sect.\ 1).  However, this link is
most likely indirect and mediated by several processes (e.g., Elmegreen 2012;
Mac Low 2013; Forbes et al.\ 2014; Goldbaum et al.\ 2015), which all together
lead to a roughly constant depletion time in the molecular gas (see chaps 9
and 10 of Krumholz 2015).

If we consider the high-redshift star-forming galaxies analysed by Genzel et
al.\ (2014), then NGC 1068 seems less unusual, at least from a qualitative
point of view.  In fact, SINS/zC-SINF galaxies at $z\sim2$ display rings and
central depressions in the inferred gas surface density, powerful AGN-driven
nuclear outflows (F\"{o}rster Schreiber et al.\ 2014), intense and widespread
bursts of star formation, as well as violent and large-scale gravitational
instabilities (see fig.\ 24 of Genzel et al.\ 2014).  In spite of such
similarities, there is an important qualitative difference: in these
galaxies, disc instabilities are driven by molecular gas and the young
generation of stars (Romeo \& Agertz 2014), rather than by the older stellar
populations.  This adds to the well-known quantitative differences that exist
between high-redshift galaxies and even their most extreme nearby analogues
(e.g., Elmegreen et al.\ 2013; Glazebrook 2013; Garland et al.\ 2015).

A fairer comparison would be between NGC 1068 and other nearby
Seyfert+starburst galaxies.  Unfortunately, to the best of our knowledge, NGC
1068 is the only galaxy of this type that has been analysed using
two-component disc instability diagnostics, i.e.\ considering not only
molecular gas but also stars.  In view of the interesting and unexpected
results found in this paper, it would be of great importance to extend our
analysis to other nearby Seyfert+starburst galaxies, and to explore the link
between gravitational instability and star formation in such galaxies.

\subsection{Highly non-trivial aspects of the problem}

Like the original Toomre parameter, $\mathcal{Q}$ measures the stability of
the disc against local, axisymmetric, linear perturbations.  Relaxing any one
of these three basic assumptions discloses a highly non-trivial aspect of the
problem (see, e.g., Binney \& Tremaine 2008; Bertin 2014).  Since we have
already discussed this issue in previous papers, here we only summarize the
main points.
\begin{enumerate}
\item The assumption of local perturbations corresponds to the condition
  $kR\gg1$, i.e.\ $\lambda\ll2\pi R$, where $k$ and $\lambda$ denote the
  radial wavenumber and the radial wavelength, respectively.  It is easy to
  check that NGC 1068 satisfies this short-wavelength approximation (see the
  bottom panels of Figs 3 and 4), and so do most spiral galaxies (see fig.\ 6
  and sect.\ 3 of Romeo \& Falstad 2013).
\item The assumption of axisymmetric, or tightly wound, perturbations is not
  so general (see sect.\ 5.2 of Romeo \& Fathi 2015).  Non-axisymmetric
  perturbations have a destabilizing effect, i.e.\ a disc with
  $1<\mathcal{Q}\leq\mathcal{Q}_{\mathrm{crit}}$ is still locally unstable
  against such perturbations.  Unfortunately, there is still no general
  consensus about the value of $\mathcal{Q}_{\mathrm{crit}}$.  For example,
  Griv \& Gedalin (2012) found that the classical estimate
  $\mathcal{Q}_{\mathrm{crit}}\approx2$ is an absolute upper limit on the
  critical stability level.  Elmegreen (2011) showed that gas dissipation has
  a similar destabilizing effect, and estimated that
  $\mathcal{Q}_{\mathrm{crit}}\approx\mbox{2--3}$.  If one assumes this local
  stability threshold, then the nearby star-forming galaxies considered in
  Sect.\ 4.1 are close to marginal instability or unstable, while NGC 1068 is
  still unusually unstable.
\item The less restrictive assumption of local non-axi\-symmetric
  perturbations is also quite controversial (see sect.\ 3 of Romeo \& Falstad
  2013).  While there is a general consensus that locally stable discs can be
  globally unstable as regards spiral structure formation, the dynamics and
  evolution of spiral structure depend critically on the radial profile of
  $\mathcal{Q}$ (see, e.g., Bertin \& Lin 1996).  Our results about
  $\mathcal{Q}(R)$ in NGC 1068 have no direct implications for that problem
  because they concern the inner $R=2\,\mbox{kpc}$, not the whole disc of NGC
  1068.
\item The assumption of linear perturbations is questionable for media with
  large-amplitude fluctuations such as the interstellar gas, where non-linear
  phenomena like turbulence and shocks are ubiquitous (see, e.g., Shu 1992).
  Non-linear effects are also important for the dynamics of bars (see, e.g.,
  Athanassoula 2013; Sellwood 2014) and for the evolution of spiral structure
  in galaxies (see, e.g., Bertin \& Lin 1996; Zhang 2012).  In other words,
  this is the most complex aspect of the problem, and is still mostly
  unsolved!  Inoue et al.\ (2016) investigated the onset and evolution of
  violent disc instabilities in high-redshift star-forming galaxies, using
  state-of-the-art simulations and diagnostics, and discussed the importance
  of non-linear effects in this process.  They showed that violent disc
  instabilities can occur even when $\mathcal{Q}$ is well above unity, and
  that non-linear effects play an important role in the formation of gas
  clumps.  Romeo et al.\ (2010) and Romeo \& Agertz (2014) showed that gas
  turbulence can excite three main instability regimes in high-redshift
  star-forming galaxies, and that two of such regimes have no classical
  counterpart: violent disc instabilities can occur at scales comparable to
  the size of gas clumps ($\sim1\,\mbox{kpc}$) even when the inferred value
  of $\mathcal{Q}$ is \emph{arbitrarily large}!  In nearby star-forming
  galaxies, the onset of disc instabilities is instead controlled by the
  value of $\mathcal{Q}$ at scales larger than $\sim100\,\mbox{pc}$ (Hoffmann
  \& Romeo 2012; Agertz et al.\ 2015).  This is clearly the case for NGC
  1068, as the BIMA SONG data have a spatial resolution of
  $\sim600\,\mbox{pc}\times400\,\mbox{pc}$ (see Sect.\ 2.1).
\end{enumerate}

\section{CONCLUSIONS}

In this paper, we have explored the role that gravitational instability plays
in NGC 1068, using robust statistics as well as reliable and predictive
diagnostics.  Our major conclusions are pointed out below.
\begin{itemize}
\item The starburst disc of NGC 1068 hosts unusually violent and large-scale
  gravitational instabilities, which are entirely driven by the self-gravity
  of the stars.  In the instability process, stars and molecular gas are
  strongly coupled, and such a coupling leads to local gravitational
  collapse/fragmentation in the molecular gas.  A comparison with other
  nearby star-forming galaxies suggests that there is a link between such
  instabilities and the unusually intense and widespread bursts of star
  formation observed in NGC 1068, although this link is most likely indirect
  and mediated by several processes.
\item Several processes, including feedback from the AGN and starburst
  activity, tend to quench disc instabilities from inside out by flattening
  down the surface density of molecular gas across the central kpc.  In the
  current evolutionary phase of NGC 1068, this is nevertheless a second-order
  effect since disc instabilities are driven by the much higher stellar
  surface density, as pointed out above.
\item Our results illustrate that even localized structures such as the
  starburst pseudo-ring and the secondary bar of NGC 1068 produce clear
  signatures in the radial profiles of the Romeo-Falstad $Q$ stability
  parameter, $\mathcal{Q}_{\mathrm{RF}}$, and characteristic instability
  wavelength, $\lambda_{\mathrm{RF}}$.  Since similar signatures have been
  detected in NGC 6946 (Romeo \& Fathi 2015) and NGC 7469 (Fathi et
  al.\ 2015), $\mathcal{Q}_{\mathrm{RF}}$ and $\lambda_{\mathrm{RF}}$ are
  promising diagnostics for probing the link between gravitational
  instabilities, starbursts and inner structures in galaxy discs, a topic
  that we will address further in future work.
\end{itemize}

\section*{ACKNOWLEDGEMENTS}

This work made use of data from: BIMA SONG, `The BIMA Survey of Nearby
Galaxies' (Helfer et al.\ 2003); SAURON, `The Spectrographic Areal Unit for
Research on Optical Nebulae' (Bacon et al.\ 2001; Emsellem et al.\ 2006); and
SDSS, `The Sloan Digital Sky Survey' (Abazajian et al.\ 2009; Bakos \&
Trujillo 2012).  We are very grateful to Oscar Agertz, Guillaume Drouart,
Suzy Jones and Lukas Lindroos for useful discussions.  We are also grateful
to an anonymous referee for constructive comments and suggestions, and for
encouraging future work on the topic.  KF acknowledges the hospitality of the
ESO Garching, where parts of this work were carried out.

\bsp

\label{lastpage}


\begin{thebibliography}{}
\bibitem{} Abazajian K. N. et al., 2009,
           ApJS, 182, 543
\bibitem{} Agertz O., Kravtsov A. V., 2015,
           preprint (arXiv:1509.00853)
\bibitem{} Agertz O., Romeo A. B., Grisdale K., 2015,
           MNRAS, 449, 2156
\bibitem{} Antonucci R. R. J., Miller J. S., 1985,
           ApJ, 297, 621
\bibitem{} Athanassoula E., 2013,
           in Falc\'{o}n-Barroso J., Knapen J. H., eds, Secular Evolution of
           Galaxies. Cambridge University Press, Cambridge, p. 305
\bibitem{} Bacon R. et al., 2001,
           MNRAS, 326, 23
\bibitem{} Bakos J., Trujillo I., 2012,
           preprint (arXiv:1204.3082)
\bibitem{} Begelman M. C., 1997,
           Ap\&SS, 248, 1
\bibitem{} Bertin G., 2014,
           Dynamics of Galaxies. Cambridge University Press, Cambridge
\bibitem{} Bertin G., Lin C. C., 1996,
           Spiral Structure in Galaxies: A Density Wave Theory. The MIT
           Press, Cambridge
\bibitem{} Bertin G., Romeo A. B., 1988,
           A\&A, 195, 105
\bibitem{} Binney J., Merrifield M., 1998,
           Galactic Astronomy. Princeton University Press, Princeton
\bibitem{} Binney J., Tremaine S., 2008,
           Galactic Dynamics. Princeton University Press, Princeton
\bibitem{} Bland-Hawthorn J., Gallimore J. F., Tacconi L. J., Brinks E., Baum
           S. A., Antonucci R. R. J., Cecil G. N., 1997,
           Ap\&SS, 248, 9
\bibitem{} Blasco-Herrera J. et al., 2010,
           MNRAS, 407, 2519
\bibitem{} Bolatto A. D., Wolfire M., Leroy A. K., 2013,
           ARA\&A, 51, 207
\bibitem{} Brinks E., Skillman E. D., Terlevich R. J., Terlevich E., 1997,
           Ap\&SS, 248, 23
\bibitem{} Bruhweiler F. C., Truong K. Q., Altner B., 1991,
           ApJ, 379, 596
\bibitem{} Cald\'{u}-Primo A., Schruba A., Walter F., Leroy A., Sandstrom K.,
           de Blok W. J. G., Ianjamasimanana R., Mogotsi K. M., 2013,
           AJ, 146, 150
\bibitem{} Daigle O., Carignan C., Amram P., Hernandez O., Chemin L.,
           Balkowski C., Kennicutt R., 2006,
           MNRAS, 367, 469
\bibitem{} Elmegreen B. G., 2011,
           ApJ, 737, 10
\bibitem{} Elmegreen B. G., 2012,
           in Tuffs R. J., Popescu C. C., eds, Proc. IAU Symp. 284, The
           Spectral Energy Distribution of Galaxies. Cambridge Univ. Press,
           Cambridge, p. 317
\bibitem{} Elmegreen B. G., Elmegreen D. M., S\'{a}nchez Almeida J.,
           Mu\~{n}oz-Tu\~{n}\'{o}n C., Dewberry J., Putko J., Teich Y.,
           Popinchalk M., 2013,
           ApJ, 774, 86
\bibitem{} Emsellem E., Fathi K., Wozniak H., Ferruit P., Mundell C. G.,
           Schinnerer E., 2006,
           MNRAS, 365, 367
\bibitem{} Fanelli M. N., Collins N., Bohlin R. C., Neff S. G., O'Connell R.
           W., Roberts M. S., Smith A. M., Stecher T. P., 1997,
           AJ, 114, 575
\bibitem{} Fathi K., 2004,
           PhD thesis, University of Groningen
\bibitem{} Fathi K., van de Ven G., Peletier R. F., Emsellem E.,
           Falc\'{o}n-Barroso J., Cappellari M., de Zeeuw T., 2005,
           MNRAS, 364, 773
\bibitem{} Fathi K. et al., 2015,
           ApJ, 806, L34
\bibitem{} Feigelson E. D., Babu G. J., 2012,
           Modern Statistical Methods for Astronomy with R Applications.
           Cambridge University Press, Cambridge
\bibitem{} Forbes J. C., Krumholz M. R., Burkert A., Dekel A., 2014,
           MNRAS, 438, 1552
\bibitem{} F\"{o}rster Schreiber N. M. et al., 2014,
           ApJ, 787, 38
\bibitem{} Gallimore J. F., Baum S. A., O'Dea C. P., Pedlar A., Brinks E.,
           1999,
           ApJ, 524, 684
\bibitem{} Garc\'{i}a-Burillo S., 2016,
           preprint (arXiv:1601.04349)
\bibitem{} Garc\'{i}a-Burillo S. et al., 2014,
           A\&A, 567, A125
\bibitem{} Garland C. A., Pisano D. J., Mac Low M.-M., Kreckel K., Rabidoux
           K., Guzm\'{a}n R., 2015,
           ApJ, 807, 134
\bibitem{} Genzel R. et al., 2014,
           ApJ, 785, 75
\bibitem{} Gerssen J., Shapiro Griffin K., 2012,
           MNRAS, 423, 2726
\bibitem{} Glazebrook K., 2013,
           PASA, 30, 56
\bibitem{} Goldbaum N. J., Krumholz M. R., Forbes J. C., 2015,
           ApJ, 814, 131
\bibitem{} Grebovi\'{c} S., 2014,
           Gravitational Instability of Nearby Galaxies: Dwarfs vs. Spirals.
           M\,Sc thesis, Chalmers University of Technology, Gothenburg,
           Sweden
\bibitem{} Griv E., Gedalin M., 2012,
           MNRAS, 422, 600
\bibitem{} Helfer T. T., Thornley M. D., Regan M. W., Wong T., Sheth K.,
           Vogel S. N., Blitz L., Bock D. C.-J., 2003,
           ApJS, 145, 259
\bibitem{} Hernandez O., Wozniak H., Carignan C., Amram P., Chemin L., Daigle
           O., 2005,
           ApJ, 632, 253
\bibitem{} Hoffmann V., Romeo A. B., 2012,
           MNRAS, 425, 1511
\bibitem{} Hopkins P. F., Torrey P., Faucher-Gigu\`{e}re C.-A., Quataert E.,
           Murray N., 2016,
           MNRAS, 458, 816
\bibitem{} Huber P. J., Ronchetti E. M., 2009,
           Robust Statistics. Wiley, Hoboken
\bibitem{} Hunter D. A., Elmegreen B. G., Rubin V. C., Ashburn A., Wright T.,
           J\'{o}zsa G. I. G., Struve C., 2013,
           AJ, 146, 92
\bibitem{} Hutchings J. B. et al., 1991,
           ApJ, 377, L25
\bibitem{} Inoue S., Dekel A., Mandelker N., Ceverino D., Bournaud F.,
           Primack J., 2016,
           MNRAS, 456, 2052
\bibitem{} Karouzos M. et al., 2014,
           ApJ, 784, 137
\bibitem{} Kennicutt R. C. Jr., 1989,
           ApJ, 344, 685
\bibitem{} Khachikian E. Y., Weedman D. W., 1974,
           ApJ, 192, 581
\bibitem{} Khoperskov S. A., Vasiliev E. O., Ladeyschikov D. A., Sobolev A.
           M., Khoperskov A. V., 2016,
           MNRAS, 455, 1782
\bibitem{} Kormendy J., Kennicutt R. C. Jr., 2004,
           ARA\&A, 42, 603
\bibitem{} Krumholz M. R., 2015,
           Notes on Star Formation. The Open Astrophysics Bookshelf
           (arXiv:1511.03457)
\bibitem{} Leroy A. K., Walter F., Brinks E., Bigiel F., de Blok W. J. G.,
           Madore B., Thornley M. D., 2008,
           AJ, 136, 2782
\bibitem{} Leroy A. K. et al., 2009,
           AJ, 137, 4670
\bibitem{} Lester D. F., Joy M., Harvey P. M., Ellis H. B. Jr., Parmar P. S.,
           1987,
           ApJ, 321, 755
\bibitem{} Mac Low M.-M., 2013,
           in Wong T., Ott J., eds, Proc. IAU Symp. 292, Molecular Gas, Dust,
           and Star Formation in Galaxies. Cambridge Univ. Press, Cambridge,
           p. 3
\bibitem{} Marcum P. M. et al., 2001,
           ApJS, 132, 129
\bibitem{} Martin C. L., Kennicutt R. C. Jr., 2001,
           ApJ, 555, 301
\bibitem{} Melioli C., de Gouveia Dal Pino E. M., 2015,
           ApJ, 812, 90
\bibitem{} Meurer G. R., Zheng Z., de Blok W. J. G., 2013,
           MNRAS, 429, 2537
\bibitem{} M\"{u}ller J. W., 2000,
           J. Res. Natl. Inst. Stand. Technol., 105, 551
\bibitem{} Neff S. G., Fanelli M. N., Roberts L. J., O'Connell R. W., Bohlin
           R., Roberts M. S., Smith A. M., Stecher T. P., 1994,
           ApJ, 430, 545
\bibitem{} Regan M. W., Thornley M. D., Helfer T. T., Sheth K., Wong T.,
           Vogel S. N., Blitz L., Bock D. C.-J., 2001,
           ApJ, 561, 218
\bibitem{} Romeo A. B., Agertz O., 2014,
           MNRAS, 442, 1230
\bibitem{} Romeo A. B., Falstad N., 2013,
           MNRAS, 433, 1389
\bibitem{} Romeo A. B., Fathi K., 2015,
           MNRAS, 451, 3107
\bibitem{} Romeo A. B., Wiegert J., 2011,
           MNRAS, 416, 1191
\bibitem{} Romeo A. B., Horellou C., Bergh J., 2004,
           MNRAS, 354, 1208
\bibitem{} Romeo A. B., Burkert A., Agertz O., 2010,
           MNRAS, 407, 1223
\bibitem{} Rousseeuw P. J., 1991,
           J. Chemometrics, 5, 1
\bibitem{} Sakamoto K., Okumura S. K., Ishizuki S., Scoville N. Z., 1999,
           ApJS, 124, 403
\bibitem{} Schinnerer E., Eckart A., Tacconi L. J., Genzel R., Downes D.,
           2000,
           ApJ, 533, 850
\bibitem{} Schoenmakers R. H. M., Franx M., de Zeeuw P. T., 1997,
           MNRAS, 292, 349
\bibitem{} Sellwood J. A., 2014,
           Rev. Mod. Phys., 86, 1
\bibitem{} Shapiro K. L., Gerssen J., van der Marel R. P., 2003,
           AJ, 126, 2707
\bibitem{} Shlosman I., Frank J., Begelman M. C., 1989,
           Nature, 338, 45
\bibitem{} Shu F. H., 1992,
           The Physics of Astrophysics Vol. 2: Gas Dynamics. University
           Science Books, Mill Valley
\bibitem{} Storchi-Bergmann T., 2014,
           in Sjouwerman L. O., Lang C. C., Ott J., eds, Proc. IAU Symp. 303,
           The Galactic Center: Feeding and Feedback in a Normal Galactic
           Nucleus. Cambridge Univ. Press, Cambridge, p. 354
\bibitem{} Telesco C. M., Decher R., 1988,
           ApJ, 334, 573
\bibitem{} Teuben P. J., 2002,
           in Athanassoula E., Bosma A., Mujica R., eds, ASP Conf. Ser. Vol.
           275, Disks of Galaxies: Kinematics, Dynamics and Perturbations.
           Astron. Soc. Pac., San Francisco, p. 217
\bibitem{} Toomre A., 1964,
           ApJ, 139, 1217
\bibitem{} Westfall K. B., Andersen D. R., Bershady M. A., Martinsson T. P.
           K., Swaters R. A., Verheijen M. A. W., 2014,
           ApJ, 785, 43
\bibitem{} Williamson D., Martel H., Kawata D., 2016,
           preprint (arXiv:1602.03817)
\bibitem{} Wong T., Blitz L., Bosma A., 2004,
           ApJ, 605, 183
\bibitem{} Yim K., Wong T., Xue R., Rand R. J., Rosolowsky E., van der Hulst
           J. M., Benjamin R., Murphy E. J., 2014,
           AJ, 148, 127
\bibitem{} Young J. S., Scoville N., 1982,
           ApJ, 258, 467
\bibitem{} Young J. S. et al., 1995,
           ApJS, 98, 219
\bibitem{} Zhang X., 2012,
           preprint (arXiv:1208.3537)
\bibitem{} Zheng Z., Meurer G. R., Heckman T. M., Thilker D. A., Zwaan M. A.,
           2013,
           MNRAS, 434, 3389
\end{thebibliography}
\end{document}